\documentclass[prl,twocolumn,showpacs,preprintnumbers,amsmath,amssymb]{revtex4} 
 
\usepackage{graphicx}
\usepackage{dcolumn}
 
\newcommand{\beq}{\begin{equation}} 
\newcommand{\eeq}{\end{equation}} 
\newcommand{\beqa}{\begin{eqnarray}} 
\newcommand{\eeqa}{\end{eqnarray}}

\begin{document} 
\title{Polaron Crossover and Bipolaronic Metal-Insulator Transition in the half-filled Holstein model.} 
 
\author{M. Capone} 
\affiliation{Enrico Fermi Center, Roma, Italy} 
\author{S. Ciuchi}  
\affiliation{Istituto Nazionale di Fisica della Materia and  
Dipartimento di Fisica\\ 
Universit\`a dell'Aquila, 
via Vetoio, I-67010 Coppito-L'Aquila, Italy}

\begin{abstract} 
The formation of a finite density polaronic state is analyzed in  
the context of the Holstein model using the Dynamical Mean-Field Theory. 
The spinless and spinful fermion cases are compared to disentangle 
the polaron crossover from the bipolaron formation. 
The  exact solution of Dynamical Mean-Field Theory is compared with  
weak-coupling  perturbation theory,  
non-crossing (Migdal), and vertex correction approximations. 
We show that polaron formation is not associated to a metal-insulator transition, which is instead due to bipolaron formation. 
\end{abstract} 
 
\pacs{71.38.-k, 71.30.+h, 71.38.Ht, 71.10.Fd} 
 
\date{\today} 
\maketitle 
 
Recent experiments strongly suggest that the  
electron-phonon (e-ph) interaction is relevant in many materials, 
ranging from high T$_c$ superconducting cuprates \cite{Lanzara},  
to colossal magnetoresistance manganites \cite{manga}, from  the  
fullerenes \cite{Gunn-review} to magnesium diboride \cite{mgb2}. 
While the specific role of the e-ph interaction is certainly very 
different from compound to compound, the properties of these materials 
are hardly explained in terms of standard approaches to the e-ph 
problem, like the Migdal-Eliashberg theory, and pose a serious 
challenge to theories. 
 
More specifically, polaronic features have been observed in  
lightly doped cuprates \cite{calvani}, in the manganites \cite{manga},  
up to some indication in the fullerenes \cite{Ricco}. 
A small polaron is a carrier so tightly interacting with the lattice 
that its effective mass is strongly enhanced therefore reducing its 
mobility\cite{notapolaroni}. The single polaron problem  
has been extensively studied allowing to understand in detail the 
polaron physics \cite{csgcdff,cgc}.  
Nevertheless, the experimental findings of polaronic effects  
obviously deal with finite densities of carriers, and  
prompt for an analogous understanding  of 
the finite density polaron problem, where the polarons are able  
to interact and to change drastically the phonon properties. 
In this regard, it is important to stress the distinction  
between polaronic and bipolaronic states. 
If repulsion between the electrons is neglected, two polarons (with 
opposite spin) tend in  fact to bind, giving rise to a (local) pair, 
which is called bipolaron. 
The bipolarons in turn may undergo a superconducting transition 
of the Bose-Einstein type \cite{alex-ranningr}. 
If superconductivity is not allowed, bipolarons   
give rise to an insulating state of localized pairs \cite{dambrumenil,bulla} 
which may eventually condense in a charge-ordered state \cite{cdp}. 
In some important compounds bipolaronic states are indeed unlikely formed. 
In cuprates and fullerenes,  the strong electron correlation forbids bipolaron 
formation, while in the manganites, the double-exchange 
mechanism favors ferromagnetic states at low temperature,  
in which no bipolaron can be formed. 
 
In order to disentangle the polaron effects from bipolaron  
formation, we compare the spinless fermion case, in which  
bipolarons are forbidden by Pauli principle, with the spinful case,  
which has been extensively studied in the recent past 
\cite{FJS+strong,freericks-weak, 
millis-shraiman,pata,millis-deppler}. 
To be more explicit, we show that the polaron crossover at finite  
density is not by itself a  metal insulator transition (MIT), and insulating 
behavior can only be associated to localized bipolarons. 
This effect shows up in the non-vanishing of both the quasiparticle 
renormalization factor, and of the renormalized phonon frequency 
in the spinless case.  
In the spinful case the quasiparticle weight vanishes at some definite 
coupling while the renormalized phonon frequency remains finite. 
 
We consider the Holstein molecular crystal model, 
in which tight-binding electrons interact with local modes of constant  
frequency. The Hamiltonian is  
\begin{equation} 
H = -t\sum_{\langle i,j\rangle,\sigma} c^{\dagger}_{i,\sigma} c_{j,\sigma} +  
H.c. - g\sum_i n_i(a_i +a^{\dagger}_i) + \omega_0 \sum_i a^{\dagger}_i a_i, 
\label{hamiltonian} 
\end{equation} 
where $c_{i,\sigma}$ ($c^{\dagger}_{i,\sigma}$) and $a_i$ ($a^{\dagger}_i$) are  
destruction (creation) operators for  fermions and for phonons of frequency $\omega_0$,  
$n_i$  the electron density, $t$ is  
the hopping amplitude, $g$ is an e-ph coupling. 
The density is fixed to $n = 0.5$ ($n=1$) in the spinless (spinful) 
system, which corresponds to the particle-hole symmetric half-filled 
case.  
In analogy with the studies of the Mott transition in the Hubbard model,
in which antiferromagnetism is neglected \cite{dmft},    
we restrict ourselves to the state with no charge order \cite{bulla}.
Despite in the particle-hole symmetric case the ground state is 
always ordered, the homogeneous solution may become representative
of the groundstate whenever charge ordering is spoiled by some
frustration effect like a next-nearest-neighbor hopping. 
Moreover, this study 
allows to characterize how strong e-ph interaction
may lead to the destruction of the metallic state (just like electron
correlation leads to the Mott state).
The adiabatic ratio $\gamma = \omega_0/t$ has been shown to be an 
important parameter for the single polaron formation.
In the adiabatic regime $\gamma <1$, the polaron crossover occurs when  
$\lambda= g^2/\omega_0 t \simeq 1$, while in the antiadiabatic  
regime $\gamma > 1$,  
the condition is instead $(g/\omega_0)^2 \simeq 1$ \cite{csgcdff,cgc}.  
 
We solve the model using the Dynamical Mean-Field Theory (DMFT),  
a non-perturbative approach which becomes exact in the limit of infinite 
dimensions \cite{dmft}.  We notice that in such a limit          
the spinless fermion case does not coincide with 
the infinite correlation limit. 
Nonetheless, this system represents an instructive 
playground where polaronic effects can be observed without too 
many competing phases. 
In DMFT, the lattice model is mapped onto an impurity 
problem subject to a self-consistency condition, which contains 
the information about the lattice. 
In our case the impurity model is 
\begin{eqnarray} 
H &=& -\sum_{k,\sigma} V_k c^{\dagger}_{k,\sigma} f_{\sigma} +  
H.c. + \sum_{k,\sigma} E_k c^{\dagger}_{k,\sigma} c_{k,\sigma} \nonumber\\ 
&-& 
g (a +a^{\dagger})\sum_{\sigma} f^\dagger_{\sigma} f_{\sigma} + \omega_0 a^{\dagger}  a, 
\label{impurity} 
\end{eqnarray} 
where the phonons live only on the impurity ($f$) site, $E_k$ and $V_k$ are
the energy levels and the hybridization parameters of
the conduction bath.  
For the $z$-coordination Bethe lattice of half-bandwidth $D=2t\sqrt{z}=1$ 
the self-consistency equation in the $z \to \infty$ limit is 
\begin{equation} 
\label{self} 
\frac{D^2}{4}G(i\omega_n) = \sum_k \frac{V_k^2}{i\omega_n - E_k}. 
\end{equation} 
In this work we use Exact Diagonalization (ED) to solve the impurity 
model (\ref{impurity}) \cite{caffarel}. This method requires to restrict 
the sum in Eqs. (\ref{impurity},\ref{self}) to a finite small number of  
levels $N_s -1$. The discretized model can then be solved at $T=0$ 
using the Lanczos method.   
The convergence is exponential in $N_s$, and   
a few levels are enough to give converged results.  
Results are obtained taking typically  $N_s = 10$,  
having checked that no significant change occurs 
for larger $N_s$ \cite{note-ph}. 
Error bars can be  evaluated by linearly extrapolating (in $1/N_s$)
to $N_s \to \infty$, and are of the size of symbols in the figures.
\begin{figure}[htbp] 
\begin{center} 
\includegraphics[width=6cm,height=2.5cm]{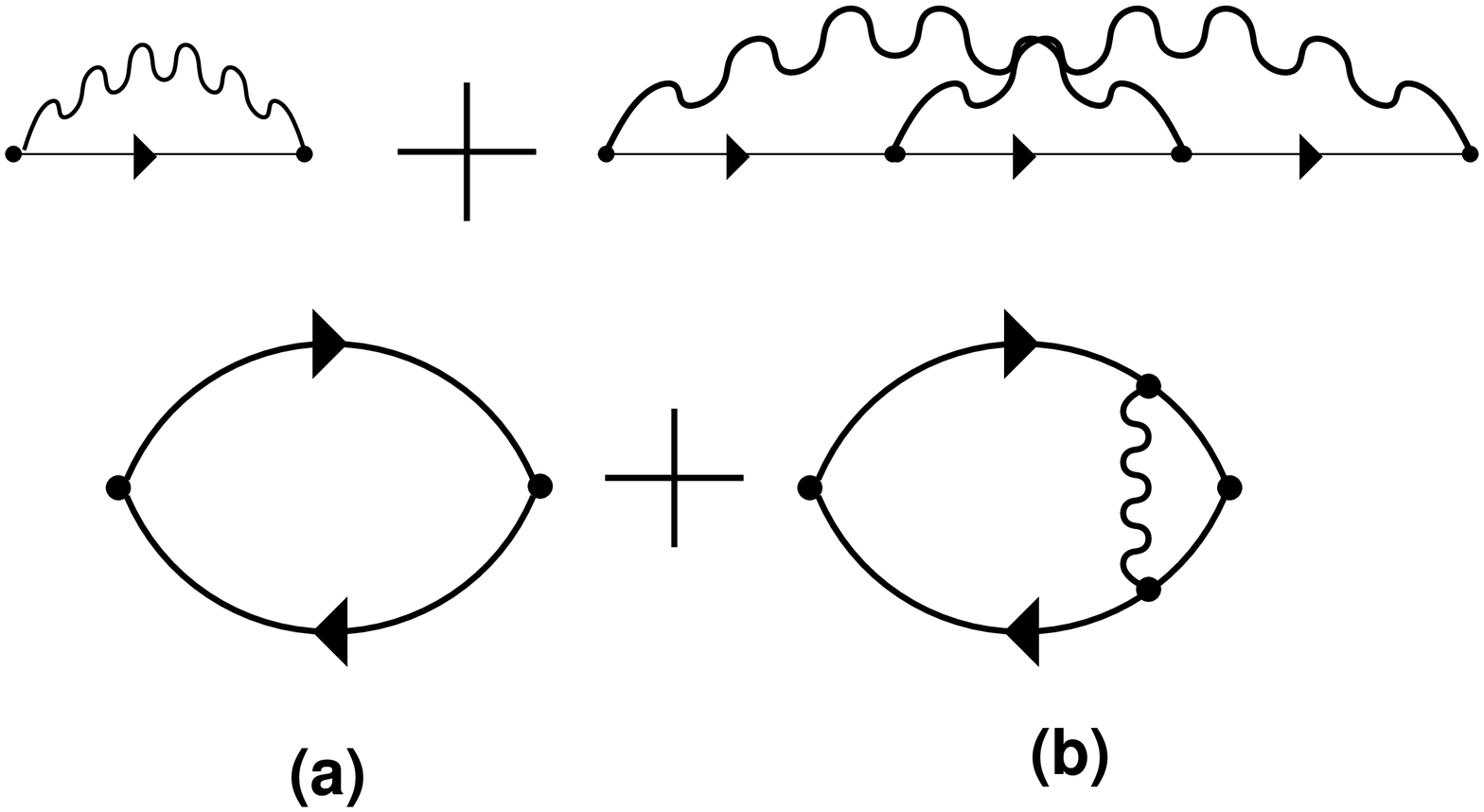} 
\end{center} 
\caption{The NCA  [(a)] and the VCA [(a) + (b)] diagrams  
for electron (upper graphs) and phonon (lower graphs) self-energies.  
Wavy lines are phonon propagators while 
straight lines are electron propagators. 
Second order perturbation theory is obtained replacing internal lines 
with free propagators.} 
\label{fig:SigmaPi} 
\end{figure}  
We compare the ED solution of DMFT, with two approximate schemes: 
a self-consistent Non Crossing Approximation (NCA) 
\cite{dambrumenil,freericks-weak} (diagrams (a) in Fig. \ref{fig:SigmaPi}) 
and a self-consistent  approximation including the first Vertex Correction  
beyond NCA (VCA) \cite{freericks-weak} (diagrams (a) and (b) in Fig.  
\ref{fig:SigmaPi}). 
Notice that in the standard Migdal-Eliashberg approximation 
the phonon spectrum is not self-consistently evaluated, but it is 
taken ``from experiments'',  
while in our NCA and VCA the phonon self-energy is self-consistently  
evaluated (on the imaginary frequency axis) through an iterative scheme, 
just like the electron self-energy.  
Actually, to allow for a safer truncation on the frequency axis,  
the quantity which is iteratively determined is not 
the full self-energy, but the difference between this quantity and the 
result from  second order perturbation theory.  
Zero temperature results are obtained by lowering the temperature  
until the physically relevant quantities converge                              
(consistently increasing the number of 
Matsubara frequecies $N_{max} = 6 / (2 T \pi)$). 
We checked the convergence to $T=0$ by monitoring  
the quasiparticle spectral weight $Z$ defined as  
$Z^{-1}= (1-(\Sigma(i\omega_{n=1})-\Sigma(i\omega_{n=0})/ 2 \pi T)$  
where $\omega_{n} = (2 n-1) \pi T$. 
The vanishing of $Z$ is used within DMFT to characterize the Mott MIT in the 
repulsive \cite{dmft}, and the pairing transition in the
attractive Hubbard models \cite{attractive}. A vanishing $Z$ has been found 
also in the spinful Holstein model \cite{bulla}.  
The convergence to $T=0$ turns out to depend on both $\gamma$ (as detailed 
below, we study $\gamma=0.1$ and $\gamma=1$ as representative of adiabatic 
and nonadiabatic regimes, respectively) and $\lambda$.  
Within VCA $T/t=3\times 10^{-3}$ is 
sufficient to get results representative of the ground state 
for weak/intermediate coupling. 
In the spinful case for $\gamma=0.1$ $T=10^{-3}$ is  
instead necessary since the polaron crossover is approached for  
smaller coupling (see below). 
Within NCA it is possible to span the strong coupling regime, making 
an extrapolation to $T=0$ necessary in the adiabatic regime 
for  $\lambda>1$ ($\lambda>0.5$)  in the spinless (spinful) case. 
In the non-adiabatic regime the extrapolation is required for 
$\lambda>1.6$ ($\lambda>1.2$) in the spinless (spinful) case.
 
\begin{figure}[htbp] 
\begin{center} 
\includegraphics[width=8cm,height=8.5cm]{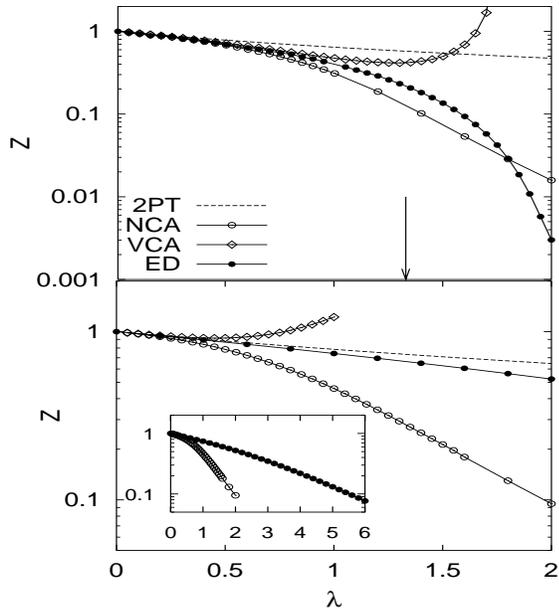} 
\end{center} 
\caption{Spinless fermions. The quasiparticle  weight $Z$ within  
second order perturbation theory (2PT),  NCA, VCA, and ED 
for $\gamma = 0.1$ (upper panel) and $\gamma = 1$ (lower panel)  
as a function of $\lambda$. The arrow in the upper panel  
marks the MIT for $\gamma=0$  \cite{millis-shraiman}. 
The inset shows ED and NCA in a larger range of $\lambda$ 
 for $\gamma=1$ to make the polaron crossover visible.} 
\label{fig:Z} 
\end{figure}

We first discuss the spinless case. 
Exact DMFT results for $Z$ are shown in Fig. \ref{fig:Z}. 
The logarithmic scale on the y-axis evidences that   
in both cases $Z$, even if exponentially reduced,  {\it does never vanish} by 
increasing $\lambda$, indicating that no MIT is taking place. It is important
to observe that $Z$ {\it increases} when the truncation error is 
reduced increasing $N_s$.
The comparison with NCA and VCA shows non trivial tendencies. 
Both  approximations are accurate at weak coupling, but VCA remains  
closer to ED for relatively large coupling even in the 
adiabatic regime, where the Migdal approximation would be expected to 
hold. Quite surprisingly even in the non-adiabatic regime  
VCA improves NCA only at weak coupling. 
At strong coupling NCA predicts a polaronic crossover, in qualitative 
agreement with exact results, even if, for $\gamma =1$, the  
crossover coupling is strongly underestimated by NCA. 
On the other hand, VCA drastically diverges from ED, and gives  
unphysical negative mass renormalization  at strong 
coupling both for $\gamma =0.1$  and $1$, in 
agreement with the divergence of vertex corrections predicted in Ref. 
\cite{dambrumenil}. The coupling at which the approximate 
methods deviate from exact results decreases with increasing $\gamma$, and 
it is unrelated with the polaron crossover coupling which is instead larger 
for $\gamma=1$, as shown in the inset of Fig. 1, and in agreement with  
the single polaron case \cite{csgcdff,cgc}. 
 
\begin{figure}[htbp] 
\begin{center} 
\includegraphics[width=8cm,height=8.5cm]{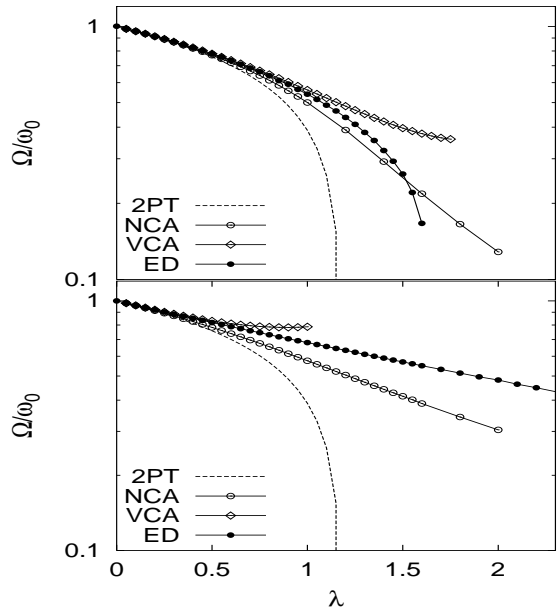} 
\end{center} 
\caption{Spinless fermions. The renormalized phonon frequency $\Omega/\omega_0$. 
Notations are the same of Fig. \ref{fig:Z}} 
\label{fig:Om} 
\end{figure}

An important difference between the finite-density situation and the 
single polaron problem is that many electrons are able to renormalize the 
phonon properties. The renormalized 
phonon frequency can be obtained from the phonon propagator   
$D(i\omega_n)= 
\int_0^{\beta} d\tau e^{i\omega_n\tau} \langle 
T(a_0(\tau)+a_0^{\dagger}(\tau))(a_0 + a_0^{\dagger})\rangle$, as 
$(\Omega/\omega_0)^2 = -2/\omega_0D(i\omega_n =0)^{-1}$.  
As  it is shown in Fig. \ref{fig:Om}, 
$\Omega/\omega_0$ never vanishes as a function of  
the coupling, but rather exponentially decreases.  
The way NCA and VCA results for $\Omega$ compare with ED is analogous  
to the results for $Z$. 
In the adiabatic regime NCA strongly overestimates $\Omega/\omega_0$ 
around the polaron crossover.  
In the nonadiabatic regime NCA again predicts a phonon softening which does 
not actually occur at such small values of the coupling.  
In the same regime VCA  improves NCA result only at weak coupling and gives 
a phonon hardening at strong coupling. 
It is worth to note that for both $Z$ and $\Omega$ 
the coupling at which NCA and VCA deviate from ED is lower in nonadiabatic  
than in adiabatic regime. 
This ``non uniform'' behavior is not present for a single polaron \cite{cgc}, 
and represents a first peculiarity of the finite-density case. 
 
Now we compare our findings for spinless fermions with 
the spinful fermion case. 
In the adiabatic regime a  MIT has been found around  
$\lambda \simeq \lambda_{MIT} \simeq 0.68$ \cite{pata}, close to the 
value ($\lambda=0.664$) at which the density of state at Fermi energy 
vanishes in the adiabatic limit ($\gamma=0$) \cite{millis-shraiman}.  
This transition has 
a precursor in the phonon softening \cite{bulla} but, contrary to the spinless 
case, the spin degrees of freedom prevent electron coherent  hopping at strong 
coupling through a Kondo-like mechanism \cite{Moeller}.  
A MIT has  been claimed to occur also within NCA \cite{dambrumenil}. 
 
\begin{figure}[htbp] 
\begin{center} 
\includegraphics[width=8cm]{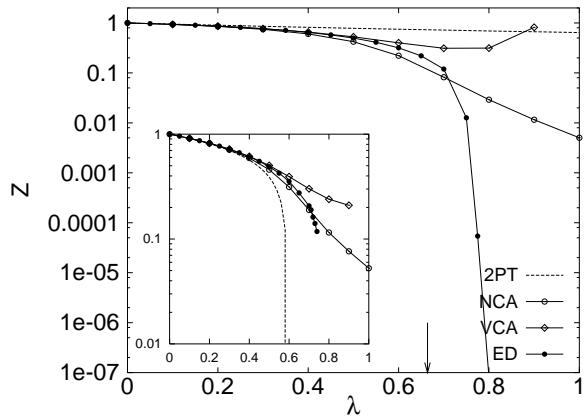} 
\end{center} 
\caption{Spinful fermions. $Z$ and  $\Omega/\omega_0$ (inset)  
for $\gamma = 0.1$ as a function of $\lambda$. 
The arrow marks the MIT for $\gamma=0$  \cite{millis-shraiman}.} 
\label{fig:spinadi} 
\end{figure}

\begin{figure}[htbp] 
\begin{center} 
\includegraphics[width=8cm]{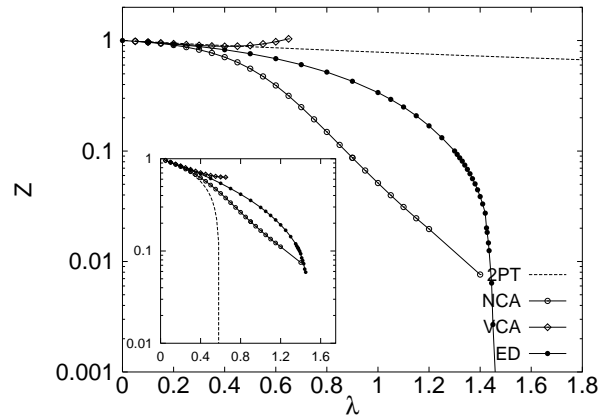} 
\end{center} 
\caption{Spinful fermions. $Z$ and  $\Omega/\omega_0$ (inset) for 
$\gamma = 1.0$ as a function of the coupling.} 
\label{fig:spinnonadi} 
\end{figure}

In Fig. \ref{fig:spinadi} we show results for $\gamma=0.1$.  
The logarithmic plot clearly shows that $Z$ vanishes 
faster than exponential, signaling a MIT 
at $\lambda \simeq 0.76$, in agreement with Ref. \cite{bulla}. Contrary to the
spinless case, here $Z$ {\it decreases} by increasing $N_s$.
On the other hand, $\Omega/\omega_0$ 
decreases with $\lambda$ as in the spinless case,  
suggesting that, even if strongly softened, the 
phonon mode never becomes completely soft. 
This behavior is more evident for larger phonon frequency $\gamma =1$,  
as reported in Fig. \ref{fig:spinnonadi}.  
Also in this case $Z$ vanishes     
much faster than exponential at $\lambda =1.44$, where the 
MIT takes place, while the phonon renormalization is much less effective 
and leads to quite a large value of $\Omega/\omega_0$ even at the MIT point 
(see inset). 

The comparison with approximate schemes qualitatively resembles the spinless case. 
NCA underestimates $Z$ for $\lambda$ below the MIT, but  
it gives a finite weight also above the exact MIT point.
Our extrapolation to $T=0$ from finite temperatures 
does not allow us to  definitively rule out the existence of a MIT within  
NCA even if, in contrast with a claim in Ref. \cite{dambrumenil},  
$Z$ seems to decrease exponentially with $\lambda$ within this 
approximation \cite{note-ambrumenil}. 
Again VCA gives better results than NCA even in the adiabatic regime 
up to intermediate coupling $\lambda\simeq 0.55$, but it becomes completely 
unreliable at strong coupling. 
 
We have studied the formation of a polaronic state in the 
Holstein model at half-filling within DMFT. For spinless    
fermions,  a continuous crossover leads to a polaronic state  
by increasing the coupling constant. Despite the electron effective mass becomes 
exponentially large in the strong-coupling regime, the ground state is always metallic.  
The crossover is more abrupt in the adiabatic case.  
In the spinful case, the polarons can bind to form bipolarons,  
leading to a real MIT.
The phonon renormalization is much stronger in the adiabatic regime than 
in the nonadiabatic case, but the phonons do not become completely 
soft at the MIT.  
Approximate treatments (NCA and VCA) strongly  
deviate from exact DMFT above a coupling 
which diminishes with increasing $\omega_0/t$. 
At weak coupling VCA correctly reproduces the qualitative 
trends of exact results, and improves on NCA.
 
M. C. thanks the hospitality and financial support of the  
Physics Department of the University of Rome ``La Sapienza'', and of  
INFM, Unit\'a Roma 1.
We acknowledge financial support of MIUR Cofin 2001 and 
enlightening discussions with C. Castellani.                   


\begin{thebibliography}{99}
\bibitem{note-ph} We also truncate the phonon Hilbert space introducing
a cut-off in the number of phonons $N_{ph}$, and
always checking for convergency.
The truncation is more delicate in the 
adiabatic regime, where a large number of phonons can be excited.  
When necessary, we used $N_{ph}=50$.
\bibitem{note-ambrumenil} The exponential decay of $Z$ within NCA shown 
in Fig. \ref{fig:spinadi} is hardly identified using a linear  
scale as in Ref. \cite{dambrumenil}.
\end{thebibliography}
\end{document}